# Selling Culture: Implementation of e-Commerce and WAP-based Prototypes

Konstantina Zafeiri[1] and Damianos Gavalas[2], Aikaterini Balla[3]

**Abstract** – *Museum stores represent integral parts of the museums that have also a lot to benefit from a successful presence on the web arena. In addition to traditional web sites, carefully designed electronic commerce (e-commerce) sites may increase the potential of museum stores offering possibilities for on-line shopping and other commercial functions. In parallel, the recent convergence of the traditionally separate technologies of the Internet and mobile telephony has brought the concept of 'wireless Internet' into the spotlight. Within this context, 'mobile commerce' (m-commerce) is a relatively new trend that represents a natural extension of e-commerce into the wireless world. M-commerce refers to electronic business transactions and differentiates from e-commerce since it involves the use of mobile devices and wireless medium rather than wired. The unique characteristics of mobile computing bring forward new challenges and opportunities for museum stores. This article presents the design and implementation of an e-commerce and an m-commerce museum shop application. The aim is to evaluate and compare the two applications in terms of several parameters, such as available technologies, strengths and limitations, design requirements, usability, interaction speed, usage cost, etc and also to identify ways for enhancing the potential of such applications and designing successful and profitable business models.*

## I. Introduction

Museums, at the most basic level, provide homes for objects. The range is vast, yet each object is linked by a conscious decision made on its behalf to be preserved, classified and displayed. Museum souvenirs solidify and materialize the experience of the visit. In turn, the museum store often features as souvenirs representative 'star exhibits' to immortalize and summarize the experience for the visitor. The visitor in the museum store, like the tourist, "is a consumer away from home" [1]. Museum shops are integral parts of modern museums and in fact have become a destination within museums. As opposed to the often monolithic, overtly didactic and somewhat threatening impression given by the museum itself, museum shops offer space for marketing and commercial exploitation. Shops' exhibits represent the contents of the museum, but in a manner that is accessible. The products of a museum shop have a different meaning and importance from any other shop product and represent an attempt to substitute cultural objects [1].

The success and popularity of the Internet and the web provides museum organizations an ideal medium for communication, documentation, promotion, advertisement and marketing. Museum shops have also a lot to benefit from a successful presence on the web ground. In addition to traditional web sites, carefully designed electronic commerce (e-commerce) sites may increase the potential of museum shops. The term e-commerce refers to any commercial exchange (delivery or transaction) of information, goods, services, and payments between entities (physical or not) over telecommunications networks [2]. Usually, e-commerce is connected with shopping and sale of information, products or services. E-commerce activities include establishing and maintaining online relationships between an organization and its suppliers, dealers, customers and other agents related to (or in support of) traditional delivery channels. Other activities include product searches and comparisons by consumers, product information presentation and promotion by manufacturers and retailers, post-purchase customer support, communication between seller and shippers or banks and other activities that are not directly related to the transaction itself.

In the context of museums, the possibilities of on-line shopping and other commercial ventures, such a ticketing and digital image libraries organization, reveal new commercial potentials for museum shops. In particular with new technologies, the space of the museum shop has extended beyond that of a physical and tangible realm. With the Internet, the world's most pervasive and gratuitous consumer marketplace, the museum shop has found itself an electronic home in several incarnations. Almost all museum web pages have links to their own private online shops and for many of that, with these online resources, it is no longer necessary to even visit the actual museum.

Beyond the "traditional" e-commerce, a new type of commerce is growing in the last decade, the mobile commerce (m-commerce). Taking into consideration the tremendous growth in mobile telephony and the evolution of the handheld devices, technologies and applications are beginning to focus more on mobile computing and the wireless web. With the 'wireless internet' becoming reality, the idea of bridging the traditional e-commerce with mobile devices has naturally shaped up. M-commerce brings forth many advantages like: ubiquity (the use of wireless device enables the user to receive information and conduct transactions anytime, anywhere); flexibility (mobile devices enable users to be contacted at virtually anytime and place); dissemination (most wireless networks support the option of synchronized information transmission to the users); personalization (customized information is enabled, meeting users preferences, followed by payment mechanisms that allow for personal information to be stored, eliminating the need to enter credit card information for each transaction) [3].

The main objective of this article is to describe an e-commerce and an m-commerce museum shop application, in order to evaluate and compare the strengths and weaknesses of the two worlds in terms of several parameters: technologies involved, limitations, design, usability, speed interaction, cost, etc. To the best of our knowledge such qualitative comparison does not exist in the literature; furthermore, although many e-commerce sites have been designed for promoting the products of museum shops, no such m-commerce application currently exists.

## II. E-Commerce Museum Shop Application

The shop of the Natural History Museum of Vrisa (Lesvos Island, Greece) has been chosen as a case study to evaluate the characteristics and requirements of museum stores e-commerce sites. The target-group of the e-shop includes experts (e.g. scientists or students of geology and paleontology), individuals interested in scientific books and the general public that wishes to learn more about the museum and its collection and purchase books, posters, cards or souvenirs.

The e-shop has been developed using osCommerce [4], an open-source platform for online e-commerce. osCommerce is a platform-independent tool that allows administrators of electronic businesses to easy install and maintain the e-shop with minimal effort and free of charge. osCommerce encompasses the following web technologies:
- the HTML (HyperText Markup Language), a markup language used to create documents on the web,
- CSS (Cascading Style Sheets), a technology that allows homogenous formatting of HTML documents,
- the PHP server-side web programming language [5], which enables the creation of web pages with dynamic content retrieved from a database,
- the MySQL Database Management System [6], which stores data related to the e-shop's products, customers, orders, etc.
- the Apache web server [7].

When installed, osCommerce creates a default e-shop with pre-defined configuration, layout, design and functionality. Post installation, the user may modify this default e-shop either through an administrator tool or though modifying the HTML/CSS and PHP code to alter or enhance the e-commerce site's functionality. The administrator tool that accompanies osCommerce platform allows the administrator to:
- regulate parameters relative to the presentation of pages,
- add, erase or modify the products categories or individual products,
- specify what kind of information is presented for each product (such as weight, size, description, manufacturer, picture, etc.),
- insert special offers for some products,
- specify the available alternatives of payment,
- administer the records of clients that have signed in as customers of the e-shop,
- administer the catalogues of orders that have been conducted by customers,
- determine the languages in which the messages will be presented (the platform supports English, German and Spanish, however support for other languages may be added),
- view reports about what products have been purchased (and in which quantity) and who are the customers that have completed purchases.

Our e-commerce application has been implemented by taking advantage of the functionality offered by the osCommerce platform along with some modifications to the appearance and the context. The most important modifications made upon the default e-shop created through the platform's installation process are the following:
- addition of new categories, products and special offers,
- offer support for displaying messages and products information in the Greek language,
- changes on the design and layout of the pages,
- changes on the appearance of the header & footer of every page,
- changes on the title of every page,
- creation of a new logo,
- changes to the existing 'chromatic pallet' (to suit to the theme of a natural history museum), the layout of the headings and the links,
- changes to the left and right column that appear on every page, as well as the boxes included within them.

The abovementioned modifications have been performed either through using an HTM/CSS authoring tool (Macromedia Dreamweaver MX) or through altering the service functionality offered by osCommerce (addition / modification of PHP code).

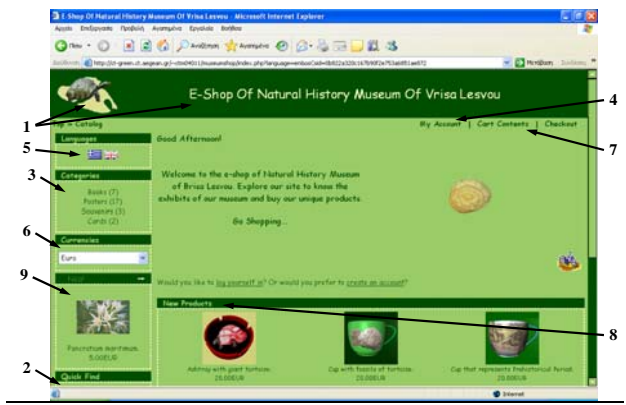

1. Logo and title of e-shop
2. Quick search field
3. Column with products categories
4. Information of account
5. Choice of language
6. Choice of currency
7. Information on the basket
8. New products, offers, reviews
9. List of new products

Figure 1: The main page of the e-shop

The products of the museum shop have been classified in four categories: books, posters, souvenirs and cards. For each product, we have entered into the platform's database information that includes the product's name (in Greek and English), price, description, thumbnail photo and full-resolution photo. Special attention has been paid on maintaining low to moderate picture sizes (up to 12 KB for thumbnails, up to 300 KB for enlarged photos) to reduce pages download times. In addition, the original layout of the left and right column has been modified. Already existing categories (boxes) have been replaced by new ones that correspond to the requirements of a museum shop. Finally, a visual effect (animation) has been incorporated on the first page, created in Macromedia Flash MX 2004. The museum e-hop is accessible at [8].

Having loaded the first page of the application, the customer notices the functionality and services offered by the museum shop. At startup, the visitor may choose a language to continue the navigation. Below the box with languages, there are the boxes with main categories, new products, advanced search, and finally the box with information about payment and delivery method (see Figure 1).

After selecting any product category, the corresponding list of products is retrieved and displayed. Having selected a product, the visitor may view the product's image, description, reviews, price, and orders that have been made from other customers. In case that the visitor desires to purchase a product, the site guides the customer until the order is completed. Some screenshot's of representative e-shop pages are shown in Figure 2.

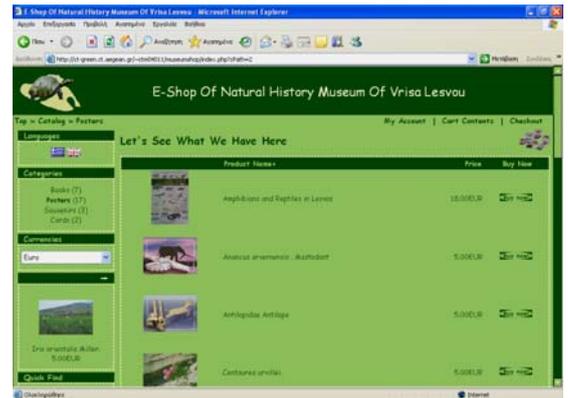
(a)

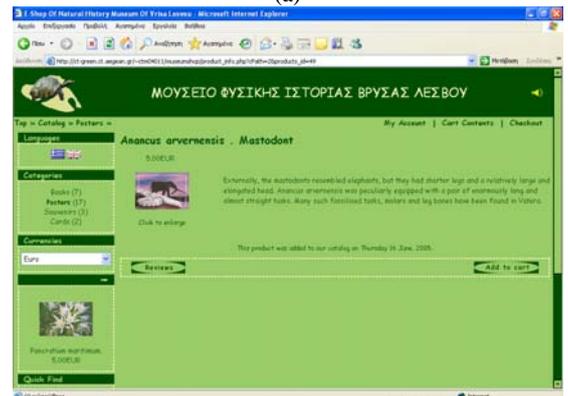
(b)

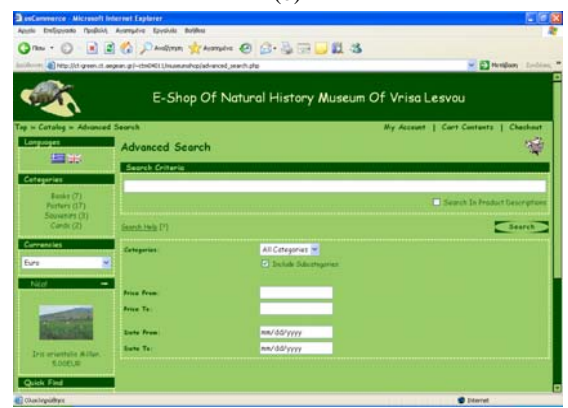
(c)

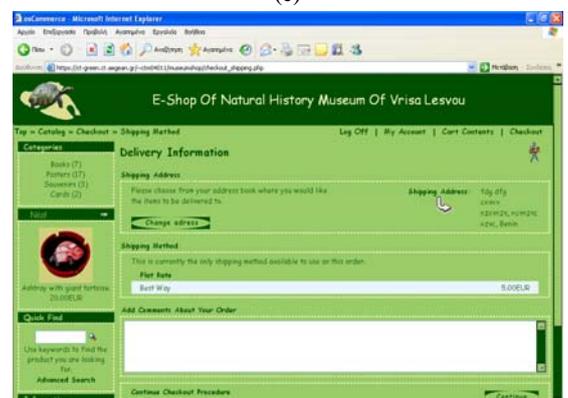
(d)

Figure 2. Screenshot from the museum shop's e-commerce application: (a) list of products for the posters category; (b) detailed information about a product, (c) products' search page; (d) delivery information for a specific order.

# III. M-Commerce Museum Shop Application

The increasingly high penetration rate of mobile phones and the consequent exposure of subscribers to mobile technology raise high expectations for the adoption of mobile commerce. Today's available mobile devices cover a broad range and include mobile phones, laptops / notebooks, Personal Digital Assistants (PDAs), palmtops, tablet PCs, pagers, etc. M-commerce (also called mobile commerce or mobile e-commerce) is defined as a special type of e-commerce still performed over the Internet, however using mobile terminals and a wireless network interface (either a wireless LAN or a network provided by a mobile operator) [3],[9].

Among others, m-commerce offers the following services [9]: information and data access (news, city guides, maps, traffic, weather, etc.); transactions (banking, brokering, shopping, auctions, betting, booking & reservation, mobile wallet); entertainment (music, games, graphics, video); communication / interaction (short messaging, unified messaging, e-mail, chat rooms, video-conferencing).

Mobile and wireless network technologies witnessed exciting innovations in recent years and will continue to represent a rapidly growing sector in the foreseeable future. The increasing demand for mobile data communications has led to the deployment of 3G mobile networks, offering higher throughput and basic multimedia services together with voice capabilities. In addition, wireless Local Area Networks (LANs) have also evolved rapidly, complementing the wireless networking landscape [10]. Mobile and wireless network technologies offer the necessary network infrastructure for the growth of m-commerce technologies and applications.

In order to guide the global development of the new wireless applications, the leaders of the telecommunications industry formed the Wireless Application Protocol Forum (WAP Forum) [11]. In particular, WAP is an open, global specification that allows users to access Internet-type content via thin-client devices, such as cell phones, pagers, PDAs, etc. The most common WAP-enabled devices are the so-called "WAP Phones". The most recently approved specification is WAP 1.2.1 (June 2000), but WAP 2.0 is now available in and is under active review and validation by the Open Mobile Alliance (OMA) [13].

The markup language used for WAP services is the Wireless Markup Language (WML) [12] which is similar to HTML. WML has been designed having resource-constrained handheld devices in mind (e.g. small displays and one-hand navigation without a keyboard or mouse). WML has a smaller set of markup tags than HTML. Unlike the flat structure of HTML documents, WML documents are divided into separate units of user interaction (termed "cards") that are easily navigable with a micro-browser.

Despite consumers' expectations, WAP has not yet met the adverts promise for a truly "mobile Internet". However, WAP has gained remarkable acceptance mainly in Europe, but also in Japan. Some of the advantages of WAP are the following:
- WAP is an open, global specification,
- WAP is designed for thin-client devices; when compared to devices used for mobile Internet services, WAP enabled devices are smaller, less cumbersome, consume less power, and are usually less expensive.

However, WAP is considered a commercial failure, which is a direct consequence of its inherent weaknesses [15]:
- most mobile devices are typically equipped with small display screens and require data entry by alphanumeric keypad strokes,
- WML supports limited text formatting and imaging,
- not many content providers offer content tailored to WAP phones, although the numbers are growing,
- WAP 1.x has been mainly operated over wireless systems wherein charges apply on connection duration basis. That has proved a major counterincentive since users were not keen on surfing the net with a tiny screen over unacceptably slow and overrated wireless connections [16].

In addition to technological issues, usability issues appear in WAP applications due to the limitations of mobile devices. The current state of wireless technology poses many constraints for designing effective user interfaces for m-commerce applications. Small screen display, limited bandwidth and the simplistic, yet, diverse functionality of wireless handheld devices certainly affect usability. Considering all the above, the following solutions have been proposed to address these problems [17],[18]:
- avoidance of scrolling, especially horizontal scrolling; scrolling, which can severely hinder users browsing behavior,
- usage of flat hierarchies (since every step takes longer on handheld devices, a flat hierarchical structure with fewer steps is preferred),
- design of a navigation system consistent with a regular web browser; this consistency enables users familiar with web browsers to transfer their browsing experience to mobile applications,
- provision of a "Back" button with the same function in a regular browser,
- provision of a history list that records the order in which hyperlinks have been traversed; the history list should present previously visited sites as a stack,
- design of an appropriate navigation system that brings users back to the data entry page after reviewing the codes, or provision of a help screen without leaving the data entry screen,
- search facilities; improvement on search precision by intelligent query support and predefined search options.

-

Regarding the design of a WAP application, the application should:
- be tested on different mobile devices, in order to ensured that the characteristics and functions are device-independent,
- incorporate small-sized graphics.

In order to gain in-depth understanding of the potentials that wireless networks offer to museum organizations, we have implemented a dynamic m-commerce application, 'parallel' to the e-commerce application introduced in Section II (i.e. the museum shop of the Natural History Museum of Vrisa has been used once again as a case-study). The application's content combines the usage of WML markup language and PHP technology. In the following paragraphs, we introduce and examine the application in terms of its implementation details, usability, services offered to the user, security, architecture, structure and data modelling aspects.

The technologies involved in the m-commerce application are the following:
- the WAP 1.1 protocol for accessing the e-shop's content through mobile terminals,
- the WML standard for formatting the e-shop's content,
- the PHP server-side programming language to enable dynamic creation of WML pages,
- the MySQL database management system to store data related to the e-shop's products, customers, orders, etc,
- the Apache web server.

The WAP site of the museum shop has been designed according to the usability guidelines listed above. The main usability features of the WAP application are:
- easy reading and browsing of information,
- multiple links that provide the user with alternative routes for locating the same information, thereby making the navigation easy and fast,
- limited number of pages with vertical scrolling,
- inclusion of a numbered list to the navigation menu items,
- small-size pages (up to 9.2 KB, including graphics),
- limited use of graphics to decrease download delays,
- inclusion of a "Back" button on every page with the same function as in regular browsers.

The main functions and characteristics of the WAP site are the following:
- provision of information for every product such as its title, description and price,
- implementation of a search engine for products search by title,
- display of the last five products added in the database of e-shop,
- user authentication (log in) page,
- implementation of a shopping cart, a personal area for users to store selected products,
- design of an order process whereby the customer, after confirming the shopping cart contents, chooses a payment method (snail mail / courier) and completes the order,
- offer of a help page and report for previous orders,
- administrator authentication (log in),
- design of pages used by the administrator to insert new products or update the information of existing ones, through a WAP phone or a desktop PC.

For the ordering process, user authentication is absolutely necessarily. The reason that this solution was preferred is that micro-browsers do not fully support cookies (a technology used in 'traditional' web interactions to identify the user that issued a request).

The interaction between the user and the WAP site goes as follows: the user of a WAP device enters the e-shop's URL (web address) in the WAP device's micro-browser. The micro-browser then posts requests to the WAP gateway. The gateway locates the site on the wired web, retrieves the requested page, encodes, compiles and forwards it to the user. The received data are then rendered for display by the mobile device's micro-browser.

Regarding data modeling, we have designed a relational MySQL database system following a similar approach to the e-commerce application. Namely the WAP site's database basically stores the same kind of data and includes information about products, product categories, customers, shopping cart contents, orders, etc.

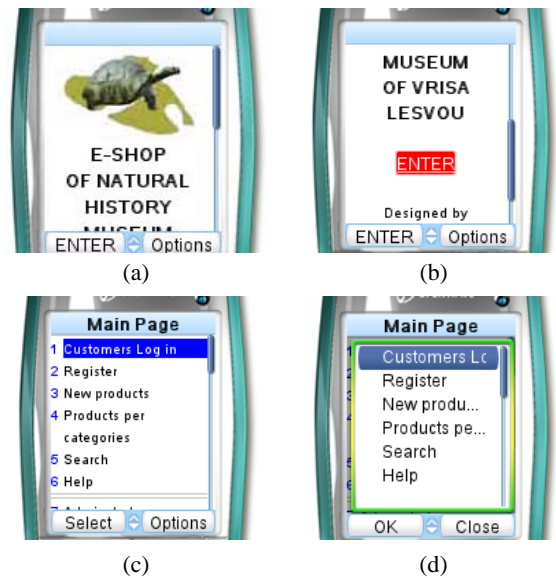

Figure 3: (a), (b) The introduction page, (c) the main page with the navigation menu, (d) selecting the 'Options' key through the keypad.

The following paragraphs describe different ways of user navigation in the WAP site of the museum shop. The Openwave Phone Simulator [14] has been used to test and debug the application. When a customer enters the WAP site, the introduction page appears that includes the logo of the museum (see Figure 3a,b). When selecting the 'ENTER' key, the customer retrieves the first page with the main navigation menu (see Figure 3c), through which he/she may choose to:

- enter as an authorized customer,
- register as a new customer to obtain username and password,
- get the last five products inserted into the database,
- browse products by category,
- search for products by title,
- ask for help.

The same navigation menu appears when user selects the "Options" key from the keyboard (see Figure 3d).

The user that wishes to enter the site as a customer is authenticated and needs to fill in his/her credentials in a login form (see Figure 4). Potential customers may sign-in through a registration page wherein they need to fill in their desired username / password pair and also their personal data (sirname, name, address, etc).

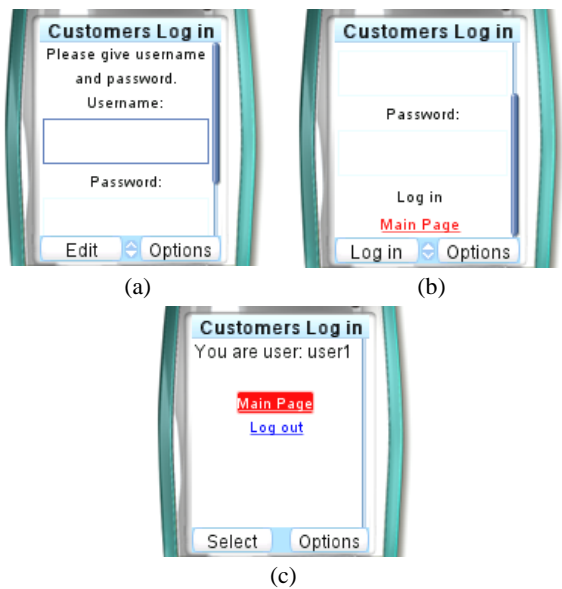

Figure 4: The log in process.

Thereafter, the user may: view the last five products added to the database (Figure 5a); view the main product categories (Figure 5b); select one of the main categories (books, posters, souvenirs or cards) to retrieve the corresponding products list (Figure 5c); obtain detailed information for a specific product (Figure 5d). Having loaded a product's card, the user may choose to add this product to his/her personal shopping cart (for customers only), or go back to the main page. If the user has not logged in, he/she is directed either to the log in or the registration page.

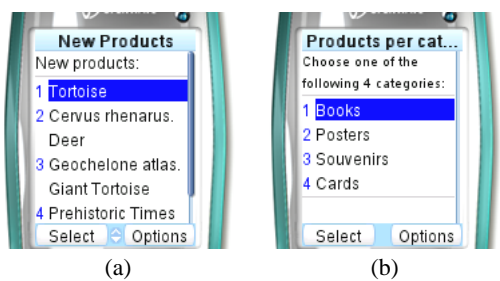

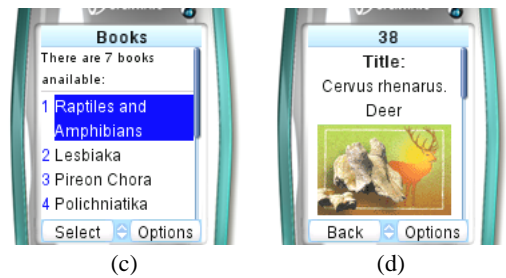

Figure 5: (a) The page with new product arrivals, (b) main product categories, (c) list of products in the 'books' category, (d) detailed information about a product.

In case of choosing to add the product to the shopping cart, a confirmation page appears (Figure 6a). At that point, the user may choose either to continue shopping or to issue an order. In the latter case, a page with the contents of the user's shopping cart appears (Figure 6b). The user may then delete a product from the cart or alter the quantity of any product. Following that, the user's personal information and preferred payment method is confirmed and the process of ordering is completed (Figure 6c).

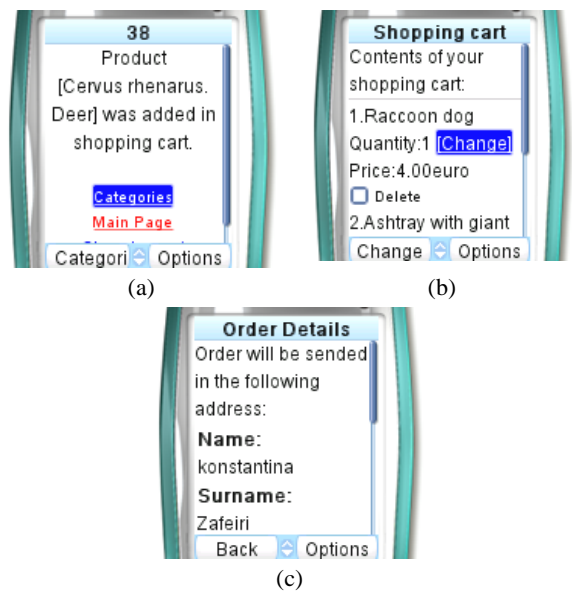

Figure 6. (a) Confirmation page of adding a product in the shopping cart; (b) the current contents of the shopping cart; (c) the order details.

The user may also choose to search a product by title. After entering a keyword to search (Figure 7a), all relevant titles are displayed (Figure 7b). Finally, the administrator(s) of the museum shop, may login into the system (Figure 8a) and load the administrator's page (Figure 8b), wherein they may choose among the following options:
- insert a new product to the database, along with the product's information;
- update the information of an existing product.

-

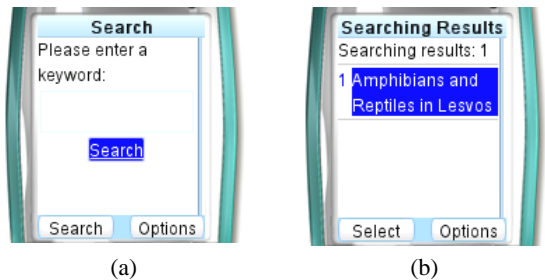

Figure 7: (a) The search page, (b) The page with searching results.

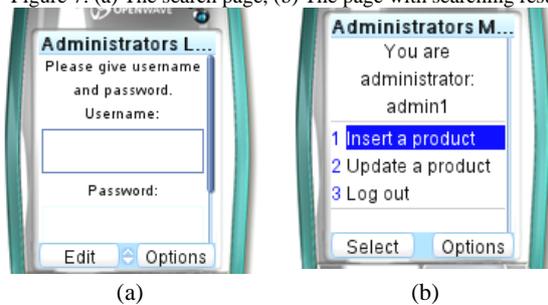

Figure 8: (a) The page with administrator's log in. (b) The main page for the administrators.

Summarizing, the design and implementation of the WAP site prototype has not been a straight-forward task. On the contrary, it proved more demanding and time-consuming compared to the e-commerce application. This is not only due to the lack of available integrated platforms (like osCommerce) that automate the development of m-commerce sites, but also due to the new programming challenges raised (e.g. lack of support for cookies) and the fact that application design requirements have been completely different thereby reducing the development speed even for experienced web developers.

## IV. Evaluation and Comparison of the E-Commerce and M-Commerce Solutions

This section aims at evaluating and comparing the two above-described applications in terms of several aspects: technologies involved, limitations, design, usability, speed interaction, cost, etc.

E-commerce applications have different characteristics from applications designed for WAP devices. The main differences are summarized below:

- Usability: Navigation and data entry in a WAP site are far more complex than in a web site [19]. PCs' keyboards are more easy to use than mobile devices' keypads. However, with the rapid penetration of mobile phones, users are getting familiar with that class of devices.
- Multimedia support: A web site usually includes a main page that may incorporate additional files (images, vides, animations, etc). In such case, the loading time of a page and the download times for external files are unacceptable for low-speed Internet connections. On the other hand, a WAP application uses one or more decks that contain a number of cards. Thus, the user can navigate from the one card to the other without changing the deck (i.e. without downloading separate files every time the user switches from a 'page' to another). In addition, small screens in WAP devices (usually up to 150×150 pixels) restrict reading and graphic's appearance in comparison with PCs' screens (usually over 800×600 pixels). Moreover, changes in environmental conditions (brightness, noise, weather, etc) may have negative affect to the users of mobile devices. Difficulty in using mobile devices leads to loss of time which frustrates users.
- Technical restrictions: WML, used in WAP applications, poses more restrictions than HTML, the de-facto standard for web applications. WML only allows pages with plain text, tables, hyperlinks, colourless images, and data entry fields. On the contrary, HTML is designed for PCs with high resolution and color depth screens, equipped with mouse and hard disks. Besides, WML is restricted in the size of WML files (up to 1.4 KB), while such restriction does not stand for HTML pages. Moreover, mobile devices' users are required to be constantly connected to a mobile / wireless phone network (constant airtime) to have access to an m-commerce application. Thus, when a user is not within a wireless network range, having access to wireless services will not be feasible either. The same restriction holds on the Internet when a user desires to have access to a site.
- Interaction speed: Most WAP devices offer speeds up to 9.6 Kbps (in 2G networks). For that reason, a WAP application is based on plain text and low-resolution graphics. On the contrary, access to the Internet through a PC requires at least a modem with speed up to 56 Kbps. Therefore, an e-commerce application can be enriched with multimedia content, without increasing download delay and cost to unacceptable levels. In addition, recent researches have shown that WAP traffic differs considerably from web traffic [20]. Hence, without modifications a web application cannpt be satisfactorily adapted to WAP. Even if page sizes decrease, the inter-arrival time will increase because of the increased reading time of every deck. Furthermore, the low potentials of mobile devices with respect to hardware discourage the incorporation of multimedia files in the applications. Even more recent mobile devices are inappropriate for the execution of demanding applications, due to the low available RAM memory (about 128-512 KB). On the contrary, the increased capabilities of PCs (~512 MB RAM for an average modern PC), allow the execution of resource-demanding applications.
- Cost: The high cost associated with accessing an application through a WAP device, prevents even simple and short navigations. This is especially true for m-commerce, where the time spent in a site is typically much longer, thereby increasing the overall cost. On the other hand, the cost for a 'wired' Internet connection is negligible, which increases the average time spent in e-commerce sites and, hence, the possibility of a customer proceeding to purchases.

## V. Conclusions & Future Work

Information and telecommunication technologies have already caused significant changes in the way museums perform their functions and the way they are perceived by the public. Initially, museum websites used to serve as digital brochures and later developed into online representations of the physical museum collections. However, this has only been the beginning. Having realized the possibilities of on-line shopping and other commercial ventures that e-commerce offers, museums have started taking advantage of the commercial and marketing opportunities offered in the web arena.

M-commerce represents a more recent type of e-commerce which has lately seen a rapid evolution, providing the ability for communication and interaction over the Internet anytime, anywhere. Despite its 'childhood illnesses', mainly related to resource constraints, interactivity, usability and communication speed, the market share of m-commerce is expected to grow even faster in the foreseeable future. Of course, the future of m-commerce will depend on a number of factors, such as: the evolution of broadband wireless networks (pricing policies, transmission rates, etc); provision of a secure framework for wireless electronic transactions; design of a new generation of mobile devices with enhanced processing capabilities, memory capacity, larger screens and simplified user input and content browsing; design of innovative services that take advantage of the unique characteristics of mobile devices. As a result, museums and museum shops have a lot to profit from a successful presence in the wireless web and from developing innovative m-commerce applications.

This article introduced an e-commerce and an m-commerce application for a natural history museum shop, as a means for evaluating and comparing their relevant strengths and weaknesses. Although many e-commerce sites for museum shops already exist, to the best of our knowledge no such m-commerce application currently exists for promoting the products of museum shops.

Regarding future work, we plan to:
- integrate additional commercial functions to both our e-commerce and m-commerce applications, such as online ticketing and digital image libraries organization,
- perform usability tests with members of our laboratory and the general public to evaluate our applications in terms of: ease of use; effectiveness and efficiency of the application; user satisfaction, etc.

## Authors information

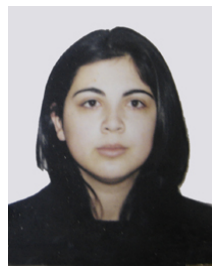
**Konstantina Zafeiri** (ctm04005@ct.aegean.gr) received her BSc degree in Informatics (Computer Science) from University of Piraeus, Greece, in 2004 and her MSc degree in "Cultural Informatics" from the Department of Cultural Technology and Communication, University of the Aegean (Greece) in 2006. She is currently a teacher of Informatics in Secondary

-

Education in Greece. Her research interests include e-commerce, m-commerce, software and IT technologies for museums and education.

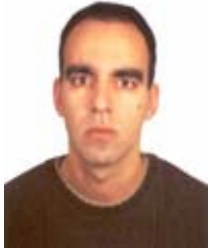

**Damianos Gavalas** (dgavalas@aegean.gr) received his BSc degree in Informatics (Computer Science) from University of Athens, Greece, in 1995 and his MSc and PhD degree in electronic engineering from University of Essex, U.K., in 1997 and 2001, respectively. He is currently an Assistant Professor in the Department of Cultural Technology and Communication, University of the Aegean, Greece. His research interests include distributed computing, mobile code, network and systems management, e-commerce, m-commerce, wireless ad-hoc and sensor networks.

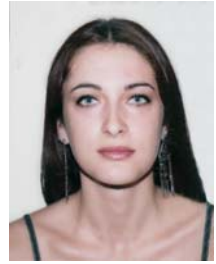

**Aikaterini Balla** (kballa@ceti.gr), received her degree in History and Archaeology from Aristotle University of Thessaloniki, Greece, in 2003 and her MSc degree in "Cultural Informatics" from the Department of Cultural Technology and Communication, University of the Aegean (Greece) in 2006. She is currently with the Cultural and Educational Technology Institute (CETI/Athena) in Xanthi, Greece. Her research interests include art history and applications of new technologies and GIS in archaeology and cultural heritage.